\documentclass[fleqn,10pt]{wlscirep}
\usepackage[utf8]{inputenc}
\usepackage[T1]{fontenc}

\title{Influence of temperature on the displacement threshold energy in graphene}

\author[1,*]{Alexandru Ionut Chirita Mihaila}
\author[1]{Toma Susi}
\author[1,*]{Jani Kotakoski}

\affil[1]{University of Vienna, Faculty of Physics, Boltzmanngasse 5, 1090 Vienna, Austria}

\affil[*]{alexandru.chirita@univie.ac.at, jani.kotakoski@univie.ac.at}

\begin{abstract}
The atomic structure of nanomaterials is often studied using transmission electron microscopy. In addition to image formation, the energetic electrons impinging on the sample may also cause damage. In a good conductor such as graphene, the damage is limited to the knock-on process caused by elastic electron-nucleus scattering. This process is determined by the kinetic energy an atom needs to be sputtered, i.e. its displacement threshold energy $E_\mathrm{d}$. This is typically assumed to have a fixed value for all electron impacts on equivalent atoms within a crystal. Here we show using density functional tight-binding simulations that the displacement threshold energy is affected by thermal perturbations of atoms from their equilibrium positions. This effect can be accounted for in the estimation of the displacement cross section by replacing the constant threshold energy value with a distribution. Our refined model better describes previous precision measurements of graphene knock-on damage, and should be considered also for other low-dimensional materials.
\end{abstract}
\begin{document}

\flushbottom
\maketitle

\thispagestyle{empty}

\section*{Introduction}

By overcoming the resolution limit of light, electron microscopes have become essential tools to investigate materials. Due to recent advances in correcting electron optical aberrations~\cite{haider_electron_1998,hawkes_p._w._aberration_2009}, transmission electron microscopes (TEMs) and scanning TEMs (STEMs) have emerged as powerful tools to provide atomic resolution~\cite{nellist_direct_2004} structural and spectroscopic information. However, the electron beam can also cause irradiation damage~\cite{kotakoski_point_2011,krasheninnikov_ion_2010}. When a fast electron approaches a nucleus, it scatters from its electrostatic potential. If momentum transfer from the electron leads to a kinetic energy of the nucleus exceeding its displacement threshold energy $E_\mathrm{d}$, a defect is created. Such knock-on collision events are the most notable irradiation effect in carbon nanostructures~\cite{banhart_irradiation_1999}. Moreover, in addition to knock-on damage, below $E_\mathrm{d}$, this process can also manipulate the structure at the level of individual atoms through bond rotations~\cite{kotakoski_stone-wales-type_2011,kotakoski_imaging_2014} and the migration of impurities in graphene~\cite{susi_manipulating_2017,tripathi_electron-beam_2018,dyck_building_2018,su_competing_2018} and crystalline silicon~\cite{jesse_direct_2018,hudak_directed_2018}. 

The meaningful parameter in experimental observations is the displacement rate, which depends on the displacement cross section $\sigma_\mathrm{d}$ of the irradiated atoms. Recent studies have shown that the atom's vibration has a great influence on its displacement, since due to momentum conservation a moving atom can gain more transferred energy than if it were static~\cite{meyer_accurate_2012,susi_isotope_2016}, resulting in significantly higher probability of damage. These studies provide the most accurate description of the displacement cross section to date, with a small discrepancy remaining between the experimental and theoretical graphene values. Despite great interest in two-dimensional materials imaging and defect formation, the description is thus still incomplete, even for the simple case of pure knock-on damage. The same formalism has been applied to materials other than graphene, including MoS$_2$~\cite{komsa_two-dimensional_2012,yoshimura_first-principles_2018} and hBN~\cite{cretu_inelastic_2015,kumar_displacement_2018}, although in non-metallic specimens inelastic excitations play an important role~\cite{cretu_inelastic_2015,lehnert_electron_2017,ievlev_building_2017}. 
Excitation lifetimes in graphene have been established~\cite{despoja_tddft_2012,yan_damping_2013,iglesias_hot_2016} to be in the order of $10^{-15}$~to~$10^{-12}$~s, while typical currents used in STEM correspond to on average one electron passing through the sample every nanosecond. Therefore, any excited states are expected to relax  between the electron impacts, and for vacancy creation in pristine graphene, only the knock-on  mechanism needs to be considered~\cite{susi_isotope_2016}.

An early electron irradiation study of copper by Jung~\cite{jung_temperature-dependence_1981} outlined a model to account for the temperature dependence of the $E_\mathrm{d}$, where he defined the $E_\mathrm{d}$ by subtracting the thermal energy contribution from the energy of a copper atom in its saddle point position. However, this effect only results in the reduction of the displacement threshold. Typically, $E_\mathrm{d}$ is treated as an intrinsic material property that describes its radiation tolerance. However, recent molecular dynamics (MD) studies~\cite{robinson_sensitivity_2012,merrill_threshold_2015} and experiments~\cite{pells_temperature_1982,zag_temperature_1983,buck_effects_1995} have shown that this is an oversimplification of the issue. For example, Merrill et al.~\cite{merrill_threshold_2015} showed the dependence of $E_\mathrm{d}$ with respect to the chirality of carbon nanotubes, while Robinson~\cite{robinson_sensitivity_2012} monitored a temperature-dependent behaviour of $E_\mathrm{d}$ in $\mathrm{TiO_2}$ and discussed the effects of temperature on the primary knock-on atom displacement threshold energies and the defect formation probability. Robinson's simulation results show a dramatic increase in the $E_\mathrm{d}$ values of O atoms, from $18\pm 3$ to $53\pm 5$ eV, at temperatures of 300 and 750 K respectively.

In this article, we prove that the displacement threshold energy $E_\mathrm{d}$ can not be assumed to be constant for each atom at finite temperatures and show that it deviates from the $E_\mathrm{d}$ value at 0~K. Importantly, we show that in contrast to the results by Robinson~\cite{robinson_sensitivity_2012}, the displacement threshold in our simulations needs to be represented as a distribution around the zero-temperature value. The width of this distribution depends on the temperature. We explain how this is due to the kinematics of the atomic lattice, using density-functional tight-binding (DFTB) simulations of graphene. We further derive an equation for calculating the maximum transferred energy resulting from the relativistic scattering of an electron from a vibrating atomic nucleus and compare it to previous approximations, and provide an improved theoretical model to predict displacement cross sections taking into account the realistic distribution of $E_\mathrm{d}$ at a given temperature $T$. Our results show that temperature needs to be taken into consideration when describing knock-on irradiation effects, not only via the atomic velocities or a simple change in the $E_\mathrm{d}$ value, but also due to the stochastic spread of $E_\mathrm{d}$ caused by thermal displacements of the atoms.

\section*{Knock-on displacements and their cross section}
Due to the great difference in mass between an electron and a nucleus (a factor of ca. 22000 for carbon), the amount of transferred energy is limited. The maximum energy is transferred in a head-on collision between the electron and nucleus, where the electron backscatters. If the transferred energy to the atom is large enough to produce a vacancy in the lattice which does not spontaneously recombine with the displaced atom, then the atom is considered displaced (knocked out). This minimum energy transfer needed to knock out an atom is called the displacement threshold energy~$E_{\mathrm{d}}$.   

In a TEM experiment with electron energies below 100 keV it is necessary to consider the vibration of the atoms in the out-of-plane direction (due to the typical experimental geometry). Resulting from the summing of the initial momenta, if an atom is hit by an electron while it happens to move parallel to the incoming electron beam it can acquire a higher transferred energy $\tilde{E_\mathrm{n}}$ than if it were at rest, as originally suggested by Brown and Augustinyak~\cite{brown_energy_1959} and later elaborated and quantified by Meyer et al.~\cite{meyer_accurate_2012} and Susi et al.~\cite{susi_isotope_2016}. To describe this situation, we need to consider a relativistic scattering process between an electron and a nucleus. The electron (mass $m$, energy $E_\mathrm{e}$, momentum $p_\mathrm{e}$), a relativistic projectile, collides with a moving non-relativistic target, the nucleus (mass $M$, energy $E_\mathrm{n}$, momentum $p_\mathrm{n}$). Taking into account energy and momentum conservation we can express the momentum of the electron after collision as

\begin{align}\label{eq:momafter}
\tilde{p}_\mathrm{e}&=\sqrt{\tilde{E_\mathrm{e}}\left(\tilde{E_\mathrm{e}}+2mc^2\right)/c^2}=\sqrt{\left(E_\mathrm{e}+E_\mathrm{n}-\tilde{E}_\mathrm{n}\right)\left(\left(E_\mathrm{e}+E_\mathrm{n}-\tilde{E}_\mathrm{n}\right)+2mc^2\right)/c^2},
\end{align}
where $c$ is the speed of light.

Given the fact that the electron kinetic energy ($E_\mathrm{e}\sim$ 80 keV) is much higher than that of the nucleus before the collision ($E_\mathrm{n}\sim 25~\mathrm{meV}$), or after  the collision ($\tilde{E}_\mathrm{n}\sim 10~\mathrm{eV}$), one can approximate $E_\mathrm{e}+E_\mathrm{n}-\tilde{E_\mathrm{n}}\approx E_\mathrm{e}-\tilde{E_\mathrm{n}}$, for which Eq.~\ref{eq:momafter} can only be solved numerically, or $E_\mathrm{e}+E_\mathrm{n}-\tilde{E_\mathrm{n}}\approx E_\mathrm{e}$, which delivers an algebraically solvable equation for the maximum energy that an electron can transfer to a nucleus moving with velocity $v$ parallel to the incident beam:

\begin{equation}\label{eq:tmaxvel}
\tilde{E}_\mathrm{n}\left(E_\mathrm{e},v\right)=\frac{\left(2\sqrt{E_\mathrm{e}\left(E_\mathrm{e}+2mc^2\right)}+Mvc\right)^2}{2Mc^2}.
\end{equation}
Setting $v=0$ would recover the result of the static nucleus approximation~\cite{banhart_irradiation_1999}.

In experimental studies, the measurable parameter is the displacement cross section. Meyer et. al.~\cite{meyer_accurate_2012} were the first to quantitatively show that under 80 keV electron irradiation, the defect-free graphene lattice remains undisturbed and that knock-on damage begins a few keV above this energy. To theoretically predict the displacement cross section, they approximated the phonon population of the material using a Debye model. The three-dimensional velocity of the atom was considered instead of the better justified one-dimensional velocity, which led to an overestimate of the out-of-plane mean square velocity by a factor of three. This model was improved upon by Susi et. al.~\cite{susi_isotope_2016} replacing the three-dimensional Debye model by the out-of-plane phonon density of states calculated with density-functional theory (DFT).

While these models both take into account the fact that the target atoms are vibrating, they assume that the displacement threshold energy itself is constant for every electron impact. In this work we present a model that includes a temperature dependency of the displacement threshold:

\begin{align}\label{eq:sigmatotal}
     &~~~~~~~~~~~~~~~~\sigma_d \left(E_\mathrm{e},v,E_{\mathrm{d}} \right)=\int_{\tilde{E}_\mathrm{n}\left(v,E_\mathrm{e}\right)\geq E_{\mathrm{d}}}P_{E_\mathrm{d}}^T\left(E\right)P^T_{v_z}\left(v \right)\sigma_d\left(\tilde{E}_\mathrm{n}\left(E_\mathrm{e}, v\right)\right)\mathrm{d}v\mathrm{d}E,
\end{align}
where $P_{E_\mathrm{d}}^T\left(E\right)$ is a normal distribution of probabilities for the displacement threshold $E_\mathrm{d}$ at a certain temperature $T$ integrated over all possible energies $E$, $P^T_{v_z}\left(v \right)$ is the probability distribution of velocities of the target atoms in the out-of-plane direction introduced in Refs.~\citenum{meyer_accurate_2012,susi_isotope_2016}, and $\sigma_\mathrm{d}$ is the displacement cross section derived by Seitz and Kohler~\cite{kinchin_displacement_1955}, which has been used to calculate cross sections for carbon nanostructures~\cite{meyer_accurate_2012,susi_isotope_2016,zobelli_electron_2007}.

\section*{Methods}

We performed density-functional tight-binding (DFTB)-based MD to study the displacement threshold energy of graphene at finite temperatures. Atomic Simulation Environment~\cite{larsen_atomic_2017} was used as the front-end for the MD to set up, perform, visualize, and analyze the results with the non self-consistent and non spin-polarized DFTB+ calculator~\cite{aradi_dftb+_2007,porezag_construction_1995,seifert_calculations_1996} (version 1.3). We employed the "matsci-0-3" parameter set in our calculations~\cite{frenzel_semi-relativistic_2004}.
Although our absolute value of the displacement threshold energy is lower than that previously obtained with DFT or experimentally estimated~\cite{susi_isotope_2016}, a similar methodology has been useful to study the dynamics of carbon atoms in graphene under electron irradiation~\cite{kotakoski_stone-wales-type_2011,krasheninnikov_stability_2005}. We point out that although more accurate DFTB-based methods are available (including self-consistent methods and spin-polarized calculations), we found that they also fail to reproduce the correct threshold but are more than an order of magnitude more demanding computationally and not necessary to establish a qualitative understanding of the variations in the threshold value depending on the temperature of the structure.
The initial structures were generated as a 15$\times$15 pristine graphene supercell structure (450~atoms) thermalized to 73, 139, 216, 307, 383 and 434~K. Displacement thresholds were estimated by picking one C atom at a time and giving it initial momentum in the $z$-direction (out-of-plane) until it got ejected (when its distance from the lattice reached at least 5~\AA), similar to earlier works~\cite{zobelli_electron_2007,kotakoski_electron_2010,kotakoski_stability_2012,susi_isotope_2016}. We repeated this process for up to 450~atoms in each thermalized structure. The time step used in the calculations was 0.3 fs and the $E_\mathrm{d}$ was estimated to within 0.1 eV for each displaced atom.

\section*{Results}

\begin{figure}[h]
\centering
    \includegraphics[width=0.82\textwidth]{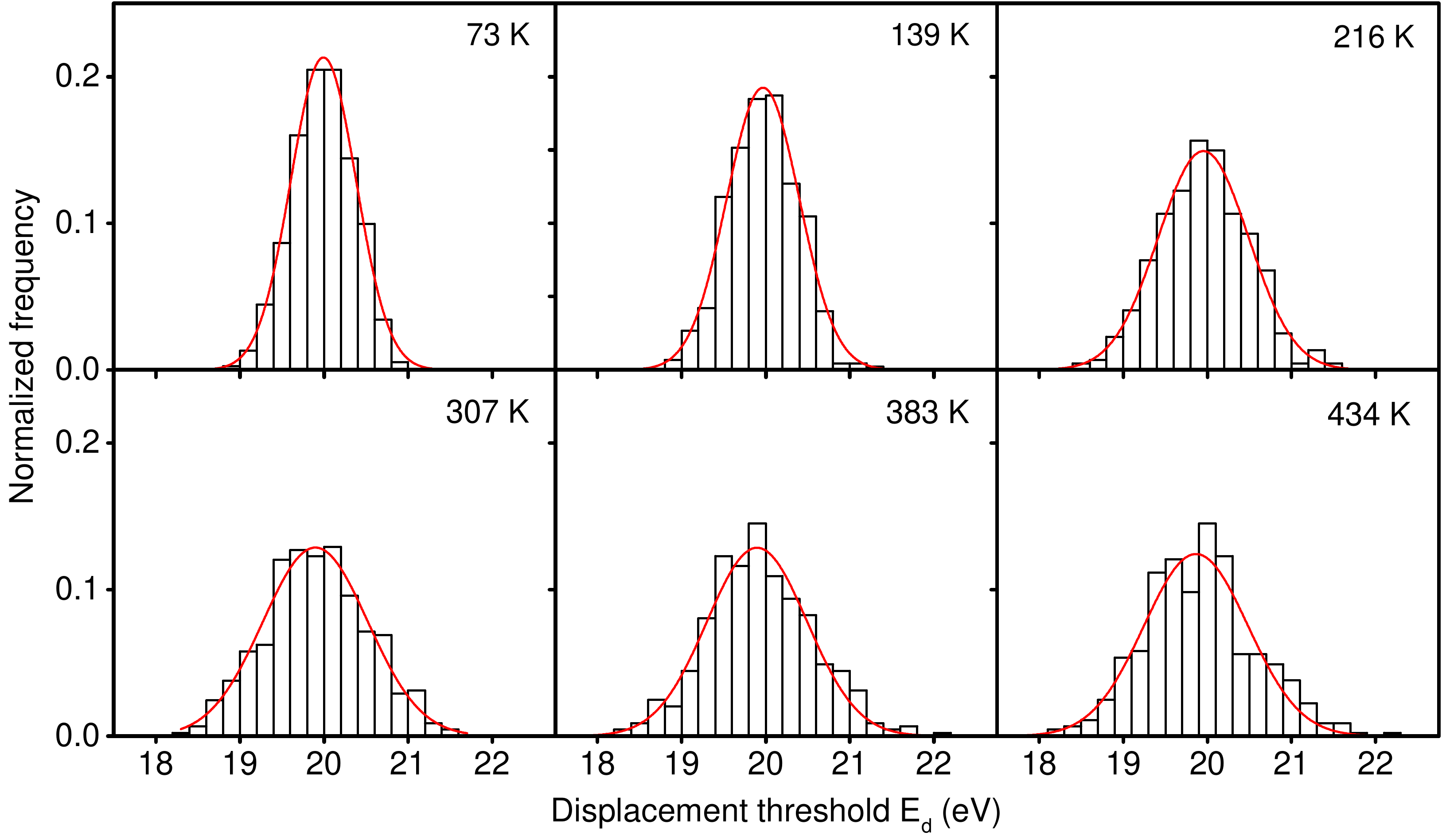}
    \caption{Histograms representing the displacement threshold distribution for structures thermalized to different temperatures. The red lines show fits of normal distributions to the data.}
    \label{fig:histodftb}
\end{figure}

\begin{figure}[h!]
\centering
    \includegraphics[width=0.6\textwidth]{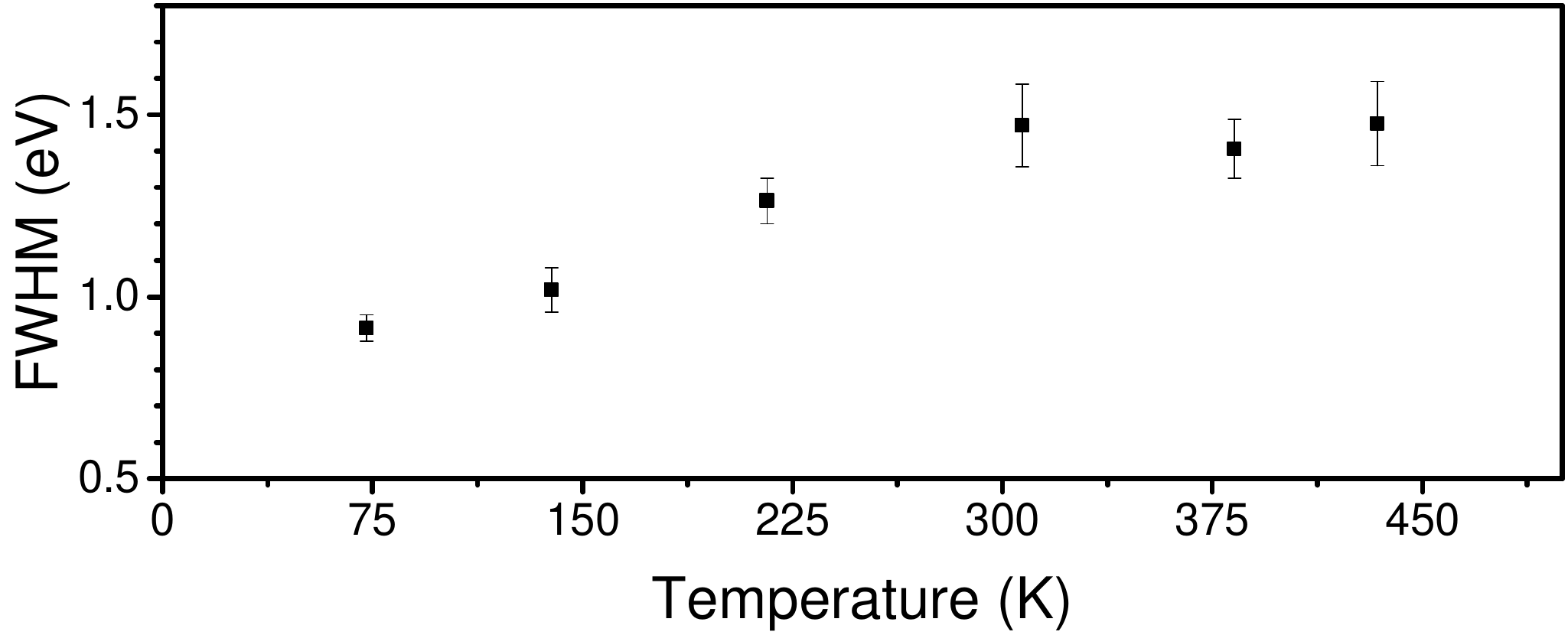}
    \caption{The FWHMs of the normal distributions resulting from the displacement simulations at different temperatures. The error bars show uncertainties of the fits of the distributions to the simulation results (Fig.~\ref{fig:histodftb}).}
    \label{fig:FWHM}
\end{figure}

The histograms in Fig.~\ref{fig:histodftb} show the spread in the displacement threshold $E_\mathrm{d}$ at different temperatures. The distribution is narrow at low temperatures and gets wider for higher temperatures. The spread is symmetric around the 0~K value of 20.0~eV, as estimated with the DFTB method, which allows us to fit a normal distribution to the histograms. In Fig.~\ref{fig:FWHM}, we plot the distribution full width at half maximum (FWHM) that varies between 0.91 and 1.47 eV depending on the temperature, starting at lower values for low temperatures and reaching a constant level at around 300~K.

\begin{figure}[h!]
\centering
    \includegraphics[width=0.7\textwidth]{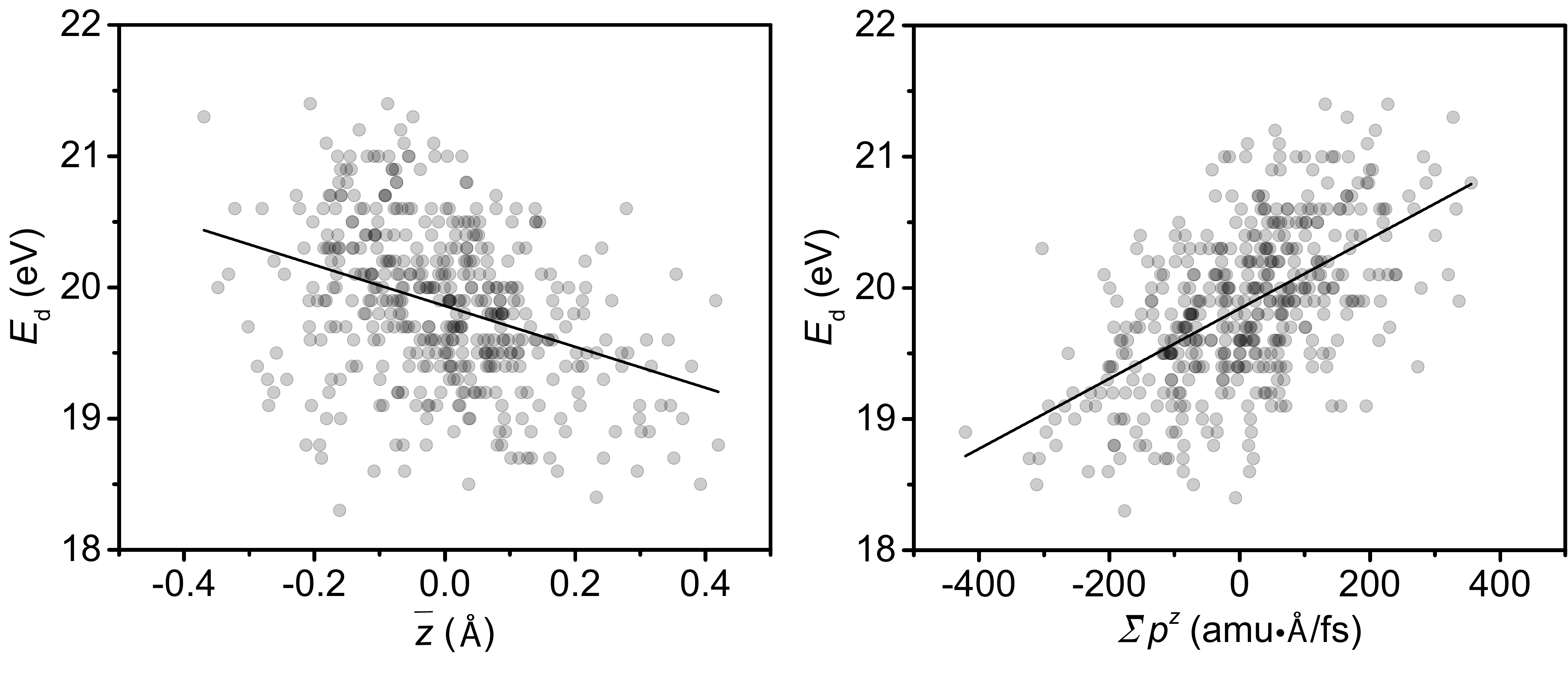}
    \caption{The correlation between the displacement threshold $E_\mathrm{d}$ at 300~K and the $z$-coordinates or the momenta of the displaced atom and its neighbors. \textbf{a}. Each point represents the average $z$-coordinate of the displaced atom and its three nearest neighbors with respect to the lattice plane at zero. \textbf{b}. Each point represents the value of the summed momenta of the displaced atom and its three nearest neighbors. The black lines represent linear fits to the data.}
    \label{fig:pos_mom}
\end{figure}

In order to understand what causes the change in the displacement threshold, we analyzed the trajectories of the simulations to gather information on the geometry and forces that might influence the outcome. We extracted information from the thermalized state (before any displacement occurred) for each atom that was about to be ejected. Our analysis suggests that there is no direct correlation between the ejecting atom's initial in-plane/out-of-plane velocity, $z$-position, or forces in- or out-of-plane and the energy~$E_\mathrm{d}$ needed to displace it. However, we do find a linear correlation between both the mean $z$-coordinate of the displaced atom $i$ and its neighbors $j=\{1,2,3\}$, $\bar{z}=\frac{1}{4}\left(z_i+\sum_{j=1}^{3}z_j\right)$ and the sum of out-of-plane momenta $\sum{p^z}=p^z_i + \sum_{j=1}^{3}p^z_j$ against the displacement threshold $E_\mathrm{d}$, as shown for 300~K in Fig.~\ref{fig:pos_mom}. As can be seen from the scatter in this data, neither one of these parameters can be used to predict the displacement threshold $E_\mathrm{d}$ for a given atom due to the complex nature of the dynamical process.

We explored the mechanism further by investigating a non-thermalized graphene sheet. Manually displacing just the ejecting atom itself in the $z$-direction with respect to its neighbours only made it easier to eject, regardless of the direction of the displacement, and thus cannot explain the observed threshold energy spread. However, if we set the initial positions of the ejecting atom and its three neighbors some tenths of an {\AA}ngstr{\"o}m above or below the rest of the atoms, we did find an asymmetric change in the displacement threshold energy. Alternatively, to confirm the connection between the neighbors and the ejecting atom, we set initial momentum values to the C atoms' nearest neighbors and started the displacement simulations. Neighbors with negative initial momenta help the C atom escape the lattice (in the positive direction) with less energy, while those with momenta towards the displacement direction contributed to a higher displacement threshold. Setting an initial momentum value for the ejecting atom itself did not explain the spread in the threshold energies, which explains our lack of success in attempting to correlate the displacement threshold only with the starting parameters of the ejecting atom itself. Hence, the ability of the nearest neighbors to follow and pull back the ejected atom, similar to what was noticed for graphene edges in Ref.~\citenum{kotakoski_stability_2012}, determines how easy it is to eject the atom.

\begin{figure}[h!]
    \centering
    \includegraphics[width=0.4\textwidth]{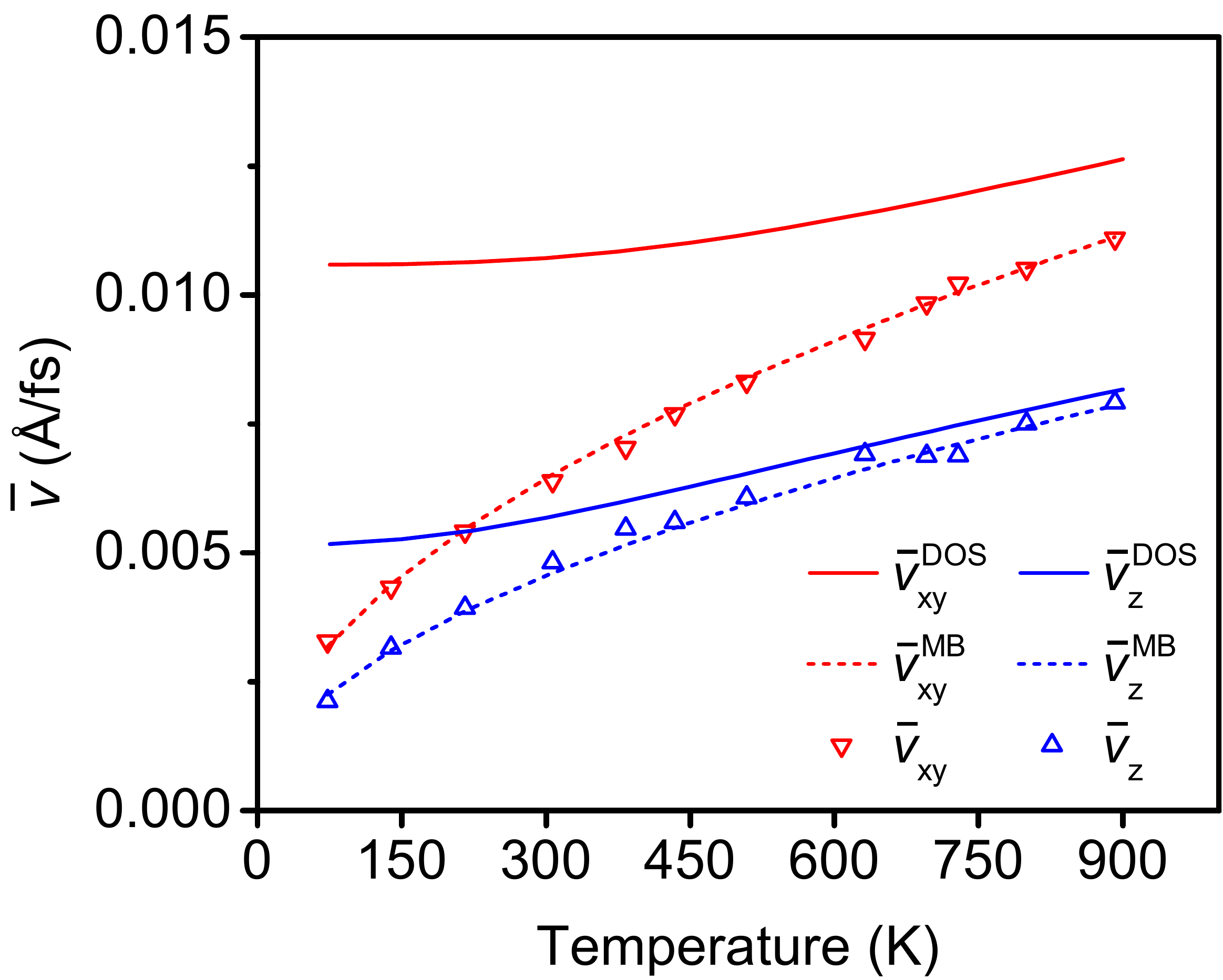}
    \caption{The root mean square (RMS) velocities of the atoms in-plane $\mathsf{\bar{v}_{xy}}$ and out-of-plane $\mathsf{\bar{v}_z}$ at different temperatures extracted from simulated trajectories represented by the triangular symbols. The linear and the dashed curves represent the RMS velocities calculated from the phonon density of states (DOS) for graphene $\mathsf{\bar{v}_{xy}^{DOS}}$, $\mathsf{\bar{v}_{z}^{DOS}}$ and the Maxwell-Boltzmann velocity distributions $\mathsf{\bar{v}_{xy}^{MB}}$ and $\mathsf{\bar{v}_{z}^{MB}}$ calculated from the ideal gas model.}
    \label{fig:velocities}
\end{figure}

Finally, to estimate the reliability of the structures thermalized via molecular dynamics within the DFTB method, we compared the root mean square velocity of the atoms for in-plane and out-of-plane directions at each temperature in Fig.~\ref{fig:velocities}. As expected, these follow the Maxwell-Boltzmann statistics. In the same plot, we show velocities calculated from the phonon density of states shown in Ref.~\citenum{susi_isotope_2016}. Clearly, the classical description within standard MD leads to an underestimation of the velocities at least up to 900~K. This means that the results we have obtained from DFTB can not be expected to give quantitatively accurate results for any given temperature. Instead, they must be considered to provide qualitative information to suggest magnitudes and trends for experiments.
 
\begin{figure}[t!]
    \centering
    \includegraphics[width=0.5\textwidth]{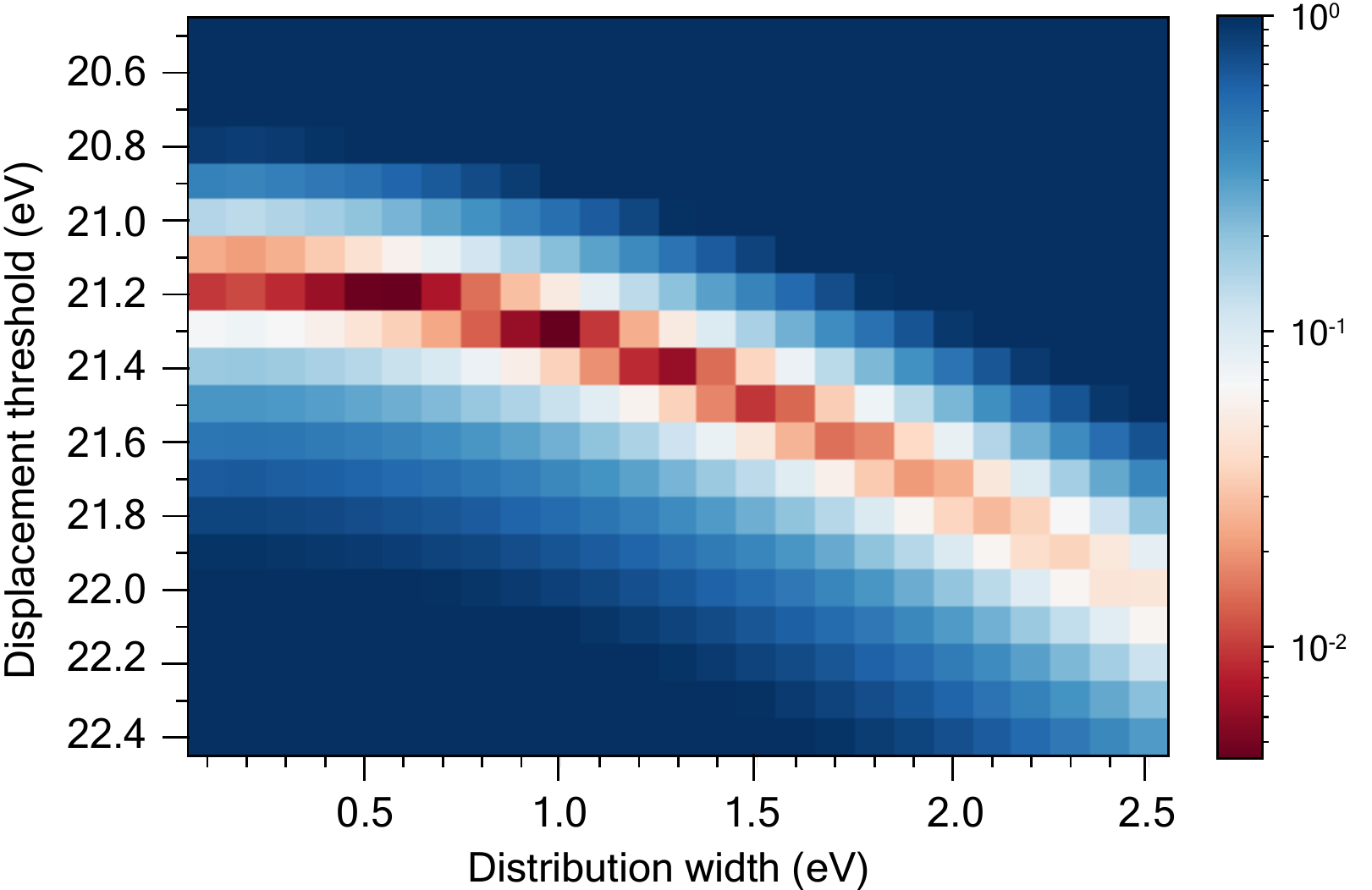}
    \caption{Heat map of the displacement cross section weighted mean squared errors at various combinations of displacement thresholds and the widths of their distribution. Two minima are present at (0.5~eV, 21.2~eV) and (1.0~eV, 21.3~eV).}
    \label{fig:tdwd}
\end{figure}
\begin{figure}[t!]
    \centering
    \includegraphics[width=0.45\textwidth]{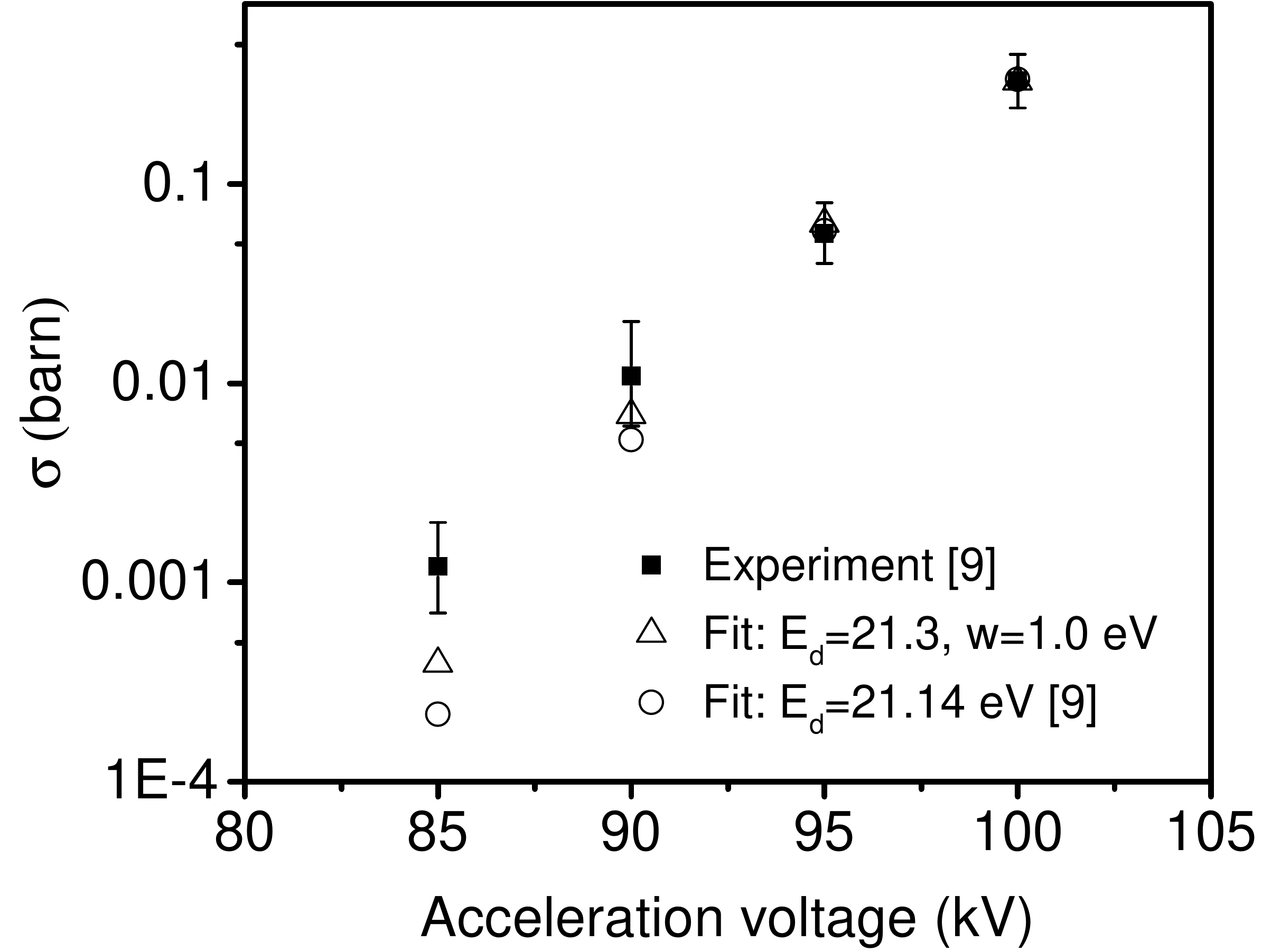}
    \caption{Comparison of fitted cross section values to STEM data from Ref.~\citenum{susi_isotope_2016}. The black squares represent the cross section values from the experiment including the uncertainties shown with error bars, while the circle and the triangle symbols correspond to the one-parameter and the two-parameter fit, respectively.}
    \label{fig:results}
\end{figure}

Finally, we compared the experimental displacement cross section data acquired with the Nion UltraSTEM 100, presented in Ref.~\citenum{susi_isotope_2016} to a two-parameter fit to Eq.~\ref{eq:sigmatotal}. We numerically integrated Eq.~\ref{eq:sigmatotal} for different displacement thresholds $E_\mathrm{d}$ and distribution widths (FWHM), and calculated the mean square error weighted by the uncertainties of the experimental values to find the best fit. We found two nearly equal minima with a relative difference of just $3.4\%$, one at $E_\mathrm{d}=21.2$~eV and $\mathrm{w}=0.5$~eV, and the other one at $E_\mathrm{d}=21.3$~eV and $\mathrm{w}=1.0$~eV (see heat map in Fig.~\ref{fig:tdwd}). In Fig.~\ref{fig:results}, we plot the experimental data and the one-parameter fit from Ref.~\citenum{susi_isotope_2016} along with our new two-parameter fit with the lowest error. Although the differences are relatively small, the new fit clearly improves upon the old values as compared to the experimental results. We point out that although this fit includes no information from the DFTB simulations, the values found for the width of the $E_d$ distribution are similar to those found in the simulations (ca. 1~eV).

\section*{Discussion}
The simulation results presented here demonstrate that the displacement threshold  energy should not be assumed to have a fixed value for low-dimensional materials when accurate estimates for the displacement cross section are needed. Simulating graphene sheets at various finite temperatures shows that the thermal perturbation of atoms from their equilibrium positions leads to a symmetric spread of their displacement threshold energies by ca. 1~eV around the 0~K value. The main reason for the spread of the displacement threshold is the variation in the momenta of the atoms neighboring the displaced atom at the moment of the impact. If these are opposite to the displacement direction, the atom will be easier to knock out. On the contrary, if their momenta are in the same direction, the atom will require more energy to knock out. We derived a formula for calculating the maximum transferred energy in the relativistic scattering of an electron at a moving nucleus, and improved the model for evaluating the theoretical displacement cross section by including the effect temperature has on the displacement threshold, providing a two-parameter fit which better describes experimental data. An accurate description of the displacements of carbon atoms from graphene is a step forward in a more precise understanding of knock-on damage also in other materials, although the magnitude is expected to be smaller in bulk materials due to the limited available space around the displaced atom.
\newpage
\bibliography{references}

\section*{Acknowledgements}
A.I.C.M. and T.S. were supported by the European Research Council (ERC) Grant 756277-ATMEN. J.K. acknowledges funding through the Austrian Science Fund (FWF) projects I3181 and P31605. We further acknowledge computational resources provided by the Vienna Scientific Cluster.

\section*{Author contributions statement}
J.K. conceived of the study. A.I.C.M. carried out the simulations and analysis of the results under the supervision of T.S. and J.K. All authors contributed to writing the manuscript.

\section*{Additional information}
\subsection*{Competing interests}
The authors declare no competing interests. 
\end{document}